\documentclass[11pt, a4paper, prd, showpacs,  preprintnumbers]{revtex4}

\usepackage{psfig, epsfig,color,amsmath,amssymb, mathrsfs}
\usepackage{graphicx}
\usepackage{fancyhdr, url, amsfonts}

\newcommand{\be}[1]{\small \begin{equation} #1 \end{equation}}

\newcommand{\abs}[1]{\left| #1 \right|}

\def\etal{{\it et al.\ }}

\begin{document}

\title{Ruppeiner  Geometry of  RN Black Holes: Flat or Curved? }

\author{B. Mirza}
\email{b.mirza@cc.iut.ac.ir}

\author{M. Zamaninasab}

\affiliation{ Department of Physics\\ Isfahan University of Technology  \\
Isfahan, 84156-83111,  Iran.}

\begin{abstract}

In some recent studies \cite{aman1, aman2, aman3}, Aman {\it et
al.} used the Ruppeiner scalar as a measure of underlying
interactions of Reissner-Nordstr\"{o}m black holes, indicating
that it is a non-interacting statistical system for which
classical thermodynamics could be used at any scale. Here, we show
that if we use the complete set of thermodynamic variables, a
non-flat state space will be produced. Furthermore, the Ruppeiner
curvature diverges at extremal limits, as it would for other types
of black holes.
\end{abstract}

\pacs{04.50.+h, 04.70.Dy}

\maketitle
\section{Introduction}
\be{\nonumber \vspace{-1.5cm}}
\par It has been commonly held
that black holes are thermodynamic systems \cite{davies, wald}.
Black holes obey four laws of black hole mechanics analogous to the
four laws of ordinary thermodynamics, posing a Bekenstein-Hawking
entropy and a characteristic Hawking temperature related to the
surface gravity of event horizon \cite{Bek,BCH,hawking1,hawking2}.
Finding the underlying microscopic description  of this entropy is
one of the most challenging subjects in theoretical physics, but it
still remains obscure, best left for future development of quantum
theory of gravity \cite{jacobson}.
\par Thermodynamic fluctuation theory, whose basic goal is to express
the time independent probability distribution for the state of a
fluctuating system in terms of thermodynamic quantities, is
usually attributed to Einstein who applied it to the problem of
blackbody radiation \cite{einstein}. However, despite a wide
range of applications, the classical fluctuation theory fails
near critical points and for volumes in the order of the
correlation volume or less. In 1979, Ruppeiner \cite{ruppeiner1}
introduced a Riemannian metric structure representing
thermodynamic fluctuation theory, and related it to the second
derivatives of the entropy.
 His theory
offered a good meaning for the distance between thermodynamic
states. He showed that the breakdown of the classical theory was due
to its failure to take  account of local correlations
\cite{ruppeiner2}. One of the most significant topics of this theory
is the introduction of the Riemannian thermodynamic curvature as a
qualitatively new tool for the study of fluctuation phenomena. This
curvature has a possible relationship with the interactions of the
underlying statistical system as proposed by Ruppeiner
\cite{ruppeiner3} in contrast with single-component ideal gas which
is a non-interacting system with zero Ruppeiner curvature. Earlier,
in 1975, Weinhold \cite{weinhold} had proposed an approach which was
based on a sort of Riemanian metric defined as the Hessian of the
internal energy of a given system, $M$, where derivatives are taken
with respect to the extensive thermodynamic variables and entropy.
 In 1984, Mruga{\l}a \cite{mrugala} and Salamon \etal
\cite{salamon} proved that these two metrics are conformally
equivalent with the inverse of the temperature, $\beta$, as the
conformal factor
\begin{eqnarray}
ds^2_R =\beta ds^2_W
\label{1}
\end{eqnarray}
   Since then, geometrical approaches have been intensively used
to study the chemical and physical properties of various
thermodynamic systems \cite{ruppeiner3}, first applied to black
holes by Ferrara {\it et al.} \cite{ferrara} to discuss the critical
behavior in moduli space. Later,  some authors used Ruppinier
geometry to study phase space, critical behavior, and stability of
various types of black hole families \cite{aman1,aman2, aman3,
cai1,cai2, sarkar}.
\par Recently, in a series of papers \cite{aman1, aman2, aman3}, Aman {\it et al.}
derived Ruppeiner and Weinhold  scalars for Reissner-Nordstr\"{o}m
(RN), Kerr, and BTZ black holes in arbitrary four dimensional
space times. One of their results is that the Ruppeiner curvature
is zero for RN black holes of any dimensions; however, for RN-AdS
and Kerr black holes thermodynamic spaces are non-trivial and
curvatures diverge at extremal limits \cite{aman1}. On the other
hand,  it
 seems that a non-interacting background can not produce the
thermodynamic  behavior of the RN black holes,  so efforts were
made to redefine Ruppeiner geometry using  new Massiue functions
\cite{cai2}. Furthermore, if one uses the Ruppeiner curvature, as
proposed in \cite{ruppeiner3}, to find a lower volume as the limit
of the applicability of classical thermodynamics, zero curvature
will tell us that there is no such limit. This seems unacceptable
because we expect classical description to fail at least at the
Planck scale.
\par  In this paper, we propose that a better measure of
microscopic interactions will be obtained if one uses the complete
phase space of extensive variables. According to this view,  in
calculating Ruppeiner curvature for RN black holes, one should set
 $J\rightarrow0, l\rightarrow\infty$ limits in the scalar
curvature of the Kerr-Newmann-AdS (KN-AdS) black hole, which is the
most general solution of Einstein-Maxwell field equations in an
anti-de Sitter background. The Ruppeiner curvature becomes a
non-zero function of $M$ and $Q$ and diverges at extremal limits.
This also indicates that classical thermodynamics could not be used
for black holes of the size of Planck length. We believe that this
is a general property which is related to the nature of fluctuation
theory and is applicable for BTZ black holes that also have  a flat
Ruppeiner curvature in Aman's method \cite{aman1}.  Furthermore,
using the quasilocal thermodynamic parameters shows that our results
do not contingent upon  using of ADM parameters.
\par  The
structure of the paper is as follows: In section II, we recall the
charged rotating solution of the Einstein-Maxwell-anti-deSitter
equations. We also recall the results of generalized Smarr formula
for this type of black holes and derive the Ruppeiner curvature.  In
section III and IV, we discuss the ideas for the RN and  Kerr
families and discuss the phase transition points and their
thermodynamic stability.  In section V, we study the Ruppeiner
geometry of KN and RN black holes by using the quasilocal
thermodynamic quantities. Throughout this paper, we use the natural
units $c=G=\hbar=1$ and also set $k_B=\frac{1}{\pi}$.
\section{Kerr-Newmann-AdS black holes}
\par Here we consider the usual charged rotating black hole in AdS
space. Its metric, which is axisymmetric, reads in Boyer-Lidquist
type coordinates \be{ ds^2 = -{\Delta_r\over\rho^2}
\left[dt-\frac{a\sin^2\theta}{\Xi}\ d\phi\right]^2
+{\rho^2\over\Delta_r}\ dr^2+{\rho^2\over\Delta_\theta}\ d\theta^2
+{\Delta_\theta\sin^2\theta\over\rho^2} \left[a\
dt-\frac{r^2+a^2}{\Xi}\ d\phi\right]^2 } where
\be{\rho^2=r^2+a^2\cos^2\theta, \qquad \Xi=1-{a^2\over l^2}}
\be{\Delta_r=(r^2+a^2) (1+{r^2\over l^2})-2mr+q^2 , \qquad
\Delta_\theta=1-{a^2\over l^2}\cos^2\theta } Here $a$ denotes the
rotational parameter, $q$ is the electric charge, and $l$  is
defined by $ l^2=\frac{-3}{\Lambda}$, where $\Lambda$ is the
cosmological constant. The mass $M$, charge $Q$ and the angular
momentum $J$ can be defined by means of Komar integrals as
\be{M=\frac{m}{\Xi^2},\qquad Q=\frac{q}\Xi, \qquad
J=\frac{am}{\Xi^2} \label{MJ}} From these relations, one can
obtain a generalized Smarr formula for Kerr-Newmann-AdS black
holes, which reads \cite{caldarelli}
\begin{widetext}
 \be{\begin{split}M=\sqrt{\frac{S}{4\pi}+\frac{\pi}{4S}(4J^2+Q^4)+\frac{Q^2}2+
\frac{J^2}{l^2}+\frac{S}{2\pi l^2}\left(Q^2+\frac{S}{\pi}
+\frac{S^2}{2\pi^2l^2}\right)}. \label{smarr2}\end{split}}
\end{widetext}
 $( S$, $J$,   $Q) $ and (\ref{smarr2}) will be regarded  as the complete set
of energetic extensive parameters and  the black hole thermodynamic
fundamental relation, $M=M(S,Q,J).$   Conjugate variables to $S$,
$Q$ and $J$ could be obtained,  for example, the Hawking temperature
is defined as : $T = \frac{\partial M}{\partial S}|_{\tiny{JQ}}$.
Extremal black holes have a zero Hawking temperature; so,
$T(S,Q,J)=0$
 is used to
 calculate the extremal surface: $J=J_{extr}(S,Q)$.
 Now, components of the Ruppeiner metric  could be derived
\be{g_{ij}=\frac{1}{T}\frac{\partial^2M}{\partial X^i\partial
X^j}\qquad X^i=(S,Q,J)}
\begin{eqnarray}
g_{11}&=&\frac{1}{2S( 9S^2+36J^2+9Q^4+18Q^2S-12J^2\Lambda
S-6S^2\Lambda Q^2-6S^3\Lambda+S^4\Lambda^2) }\nonumber\\
&\times& \frac{1}{ \left(
3{S}^{2}-12{J}^{2}-3{Q}^{4}-2{S}^{2}\Lambda{Q}^{2}-4{S}^{3}\Lambda+{S}^{4}{\Lambda}^{2}
\right)}
\times\bigl(-27{S}^{4}-432{J}^{2}{S}^{2}\Lambda{Q}^{2}\nonumber\\
&-&108{S}^{4}\Lambda{Q}^{2}
-864{J}^{2}{S}^{3}\Lambda+360{J}^{2}{S}^{4}{\Lambda}^{2}
+1296{J}^{4}+81{Q}^{8}
+648{S}^{2}{J}^{2}\nonumber\\
&+&162{S}^{2}{Q}^{4} +18{S}^{6}{\Lambda}^{2}
+648{J}^{2}{Q}^{4}-8{S}^{7}{\Lambda}^{3}
+{S}^{8}{\Lambda}^{4}-108{Q}^{6}{S}^{2}\Lambda\nonumber\\
&-&216{Q}^{4}{S}^{3}\Lambda+54{Q}^{4}{S}^{4}{\Lambda}^{2}
+72{S}^{5}{\Lambda}^{2}{Q}^{2}
-12{S}^{6}{\Lambda}^{3}{Q}^{2}-144{Q}^{4}{J}^{2}\Lambda
S\nonumber\\
&+&216{Q}^{6}S +864{J}^{2}{Q}^{2}S-576{J}^{4}\Lambda
S-48{S}^{5}{\Lambda}^{3}{J}^{2}\bigr) \end{eqnarray}
\begin{eqnarray}
g_{12} =g_{21}&=& \frac{2Q}{\left(
9{S}^{2}+36{J}^{2}+9{Q}^{4}+18{Q}^{2}S-12{J}^{2}\Lambda
S-6{S}^{2}\Lambda{Q}^{2}-6{S}^{3}\Lambda+{S}^{4}{\Lambda}^{2}
\right)} \nonumber\\
&\times& \frac{1}{ \left(
3{S}^{2}-12{J}^{2}-3{Q}^{4}-2{S}^{2}\Lambda{Q}^{2}-4{S}^{3}\Lambda+{S}^{4}{\Lambda}^{2}
\right) } \times  \bigl(
-27{S}^{3}+{S}^{6}{\Lambda}^{3}\nonumber\\
&+&108S{J}^{2} -81{Q}^{2}{S}^{2}+54{Q}^{2}{S}^{3}\Lambda
-9{Q}^{2}{S}^{4}{\Lambda}^{2}-108{J}^{2}{Q}^{2}
-27{Q}^{6}\nonumber\\
&-&108{J}^{2}{S}^{2}\Lambda
+27{Q}^{4}{S}^{2}\Lambda+27{S}^{4}\Lambda
-81{Q}^{4}S-9{S}^{5}{\Lambda}^{2}\nonumber\\
&+&72{Q}^{2}{J}^{2}\Lambda S+24{S}^{3}{\Lambda}^{2}{J}^{2} \bigr)
\end{eqnarray}
\begin{eqnarray}
g_{13}=g_{31}&=&\frac {12J }{ \left(
9{S}^{2}+36{J}^{2}+9{Q}^{4}+18{Q}^{2}S-12{J}^{2}\Lambda
S-6{S}^{2}\Lambda{Q}^{2}-6{S}^{3}\Lambda+{S}^{4}{\Lambda}^{2}
\right)} \nonumber\\
&\times&\frac{1}{\bigl(
3{S}^{2}-12{J}^{2}-3{Q}^{4}-2{S}^{2}\Lambda{Q}^{2}-4{S}^{3}\Lambda+{S}^{4}{\Lambda}^{2}
\bigr)}\times
\bigl(-27{S}^{2}+27{S}^{3}\Lambda\nonumber\\
&-&36{J}^{2}+12{J}^{2}\Lambda
S-9{Q}^{4}-3{Q}^{4}S\Lambda+18{S}^{2}\Lambda{Q}^{2}-2{S}^{3}{\Lambda}^{2}{Q}^{2}-9{S}^{4}{\Lambda}^{2}\nonumber\\
&+&{S}^{5} {\Lambda}^{3}-36{Q}^{2}S \bigr)
\end{eqnarray}
\begin{eqnarray}
g_{22}&=& -\frac{4S}{\left(
9{S}^{2}+36{J}^{2}+9{Q}^{4}+18{Q}^{2}S-12{J}^{2}\Lambda
S-6{S}^{2}\Lambda{Q}^{2}-6{S}^{3}\Lambda+{S}^{4}{\Lambda}^{2}
\right)}\nonumber\\
&\times&\frac{ 1 }{\left(
3\,{S}^{2}-12\,{J}^{2}-3\,{Q}^{4}-2\,{S}^{2}\Lambda\,{Q}^{2}-4\,{S}^{3}\Lambda+{S}^{4}{\Lambda}^{2}
\right) }\times\bigl(
-27{S}^{3}+{S}^{6}{\Lambda}^{3}\nonumber\\
&-&108S{J}^{2}-81{Q}^{2}{S}^{2}+54{Q}^{2}{S}^{3}\Lambda-9{Q}^{2}{S}^{4}{\Lambda}^{2}
-324{J}^{2}{Q}^{2}-27{Q}^{6}\nonumber\\
&+&72{J}^{2}{S}^{2}\Lambda+27{Q}^{4}{S}^{2}\Lambda+27{S}^{4}\Lambda
-81{Q}^{4}S-9{S}^{5}{\Lambda}^{2}\nonumber\\
&+&108{Q}^{2}{J}^{2}\Lambda S-12\,{S}^{3}{\Lambda}^{2}{J}^{2} \bigr)
\end{eqnarray}
\begin{eqnarray}
g_{23}=g_{32}&=&-\,{\frac {48SQJ \left(
-3\,{Q}^{2}-3\,S+{S}^{2}\Lambda \right)  \left( -3+S\Lambda \right)
}{ \left(
9\,{S}^{2}+36\,{J}^{2}+9\,{Q}^{4}+18\,{Q}^{2}S-12\,{J}^{2}\Lambda\,S-6\,{S}^{2}\Lambda\,{Q}^{2}-6\,{S}^{3}\Lambda+{S}^{4}{\Lambda}^{2}
\right)   }}\nonumber\\
&\times&\frac{1}{\left(
3\,{S}^{2}-12\,{J}^{2}-3\,{Q}^{4}-2\,{S}^{2}\Lambda\,{Q}^{2}-4\,{S}^{3}\Lambda+{S}^{4}{\Lambda}^{2}
\right)}
\end{eqnarray}
\begin{eqnarray}
g_{33}&=& -{\frac {8S   \left(
9\,{S}^{2}-6\,{S}^{2}\Lambda\,{Q}^{2}+9\,{Q}^{4}+18\,{Q}^{2}S-6\,{S}^{3}\Lambda+{S}^{4}{\Lambda}^{2}
\right)\left( -3+S\Lambda \right) }{ \left(
9\,{S}^{2}+36\,{J}^{2}+9\,{Q}^{4}+18\,{Q}^{2}S-12\,{J}^{2}\Lambda\,S-6\,{S}^{2}\Lambda\,{Q}^{2}-6\,{S}^{3}\Lambda+{S}^{4}{\Lambda}^{2}
\right)   }}\nonumber\\
&\times&\frac{1}{\left(
3\,{S}^{2}-12\,{J}^{2}-3\,{Q}^{4}-2\,{S}^{2}\Lambda\,{Q}^{2}-4\,{S}^{3}\Lambda+{S}^{4}{\Lambda}^{2}
\right)}
\end{eqnarray} Furthermore, by definition of the Ricci scalar  \cite{carroll}, the Ruppeiner curvature could be
calculated directly by using GRTensor II or GRTensor M packages.
 The expression is too
complicated to present here, so the results are depicted
numerically. The Ruppeiner curvature diverges at the extremal limit,
$J_1= J_{extr}(S,Q)$, and along the surface $J_2(S,Q)$ where metric
changes its sign (a signature of thermal instability). For different
combinations of $S$ and $Q$, each $J$ function has a double root, so
at four physically acceptable values of  $J$, the scalar curvature
diverges. The Ruppeiner geometry is not flat (Fig. 1), a fact that
could also be checked by the Cotton-York tensor.
\begin{figure}
\centering \psfig{file=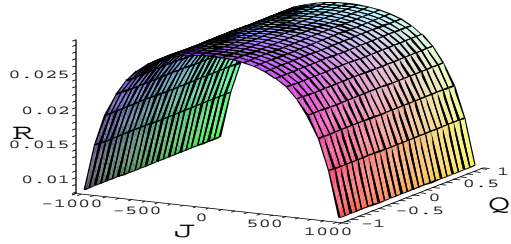, width=.4\textwidth, height=4cm}
\caption{ {\footnotesize  Ruppeiner curvature of KN-AdS BHs as a
function of $Q$ and $J$.  We set $S=100$ and  $\Lambda=-1$. The
plot shows  the region between the first two divergent points,
which indicates that it is a non-flat Ruppeiner curvature. }}
\label{fig: 1 }
\end{figure}

\section{RN black holes, the matter of  statistical interactions?}
 Up untill now, we have been considering the RN case which is our
main interest, because of the works by Aman {\it et al.}
\cite{aman1,aman2,aman3},  indicating that this family of black
holes has a trivial thermodynamic geometry (zero curvature) and so
has a non-interacting underlying statistical system. This can be
an interesting result which may serve as a guide in looking for an
appropriate statistical model for black holes in loop quantum
gravity or  string theory.  On the other hand, it seems that the
physical  structure of RN black holes could not be reproduced by
such a simple underlying system.  Furthermore,  the absence of the
divergent points set it in a different class from other types of
black holes. If  RN   systems are viewed as the limit of KN-AdS
black holes by setting $l\rightarrow\infty$ and $J\rightarrow0$ in
the Ruppeiner curvature calculated for KN-AdS case, $R$ will
takes the form
\be{R=\frac{S^2+Q^2S+2Q^4}{(S^2-Q^4)(S+Q^2)}} \\
 We now have a new
non-zero Ruppeiner scalar which diverges at the extremal limits:
$Q=\pm \sqrt{S}$ (Fig. 2). This observation is in agreement with
\cite{pavon}, which used the second moments of the fluctuations in
the fluxes of energy and angular momentum. It was shown that phase
transitions  occur only at the extremal limits, the points at
which a  black hole changes its nature to a naked singularity, a
new phase.
\begin{figure}
\centering \psfig{file=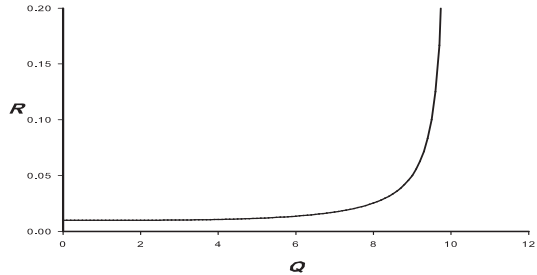, width=.4\textwidth, height=4cm}
\caption{ {\footnotesize Ruppeiner scalar of RN BH
($l\rightarrow\infty, J\rightarrow0$) as a function of $Q$. $R$
diverges   at $Q=\pm10$. Even at $Q=0$, $R$ has a $\frac{1}{S}$
remnant.  } } \label{fig: 2 }
\end{figure}
\par It should be noted that this  behavior is related to the nature of
the Ruppeiner geometry. If we initially work with  spinless black
holes
\begin{eqnarray}
g_{ij}&=&\frac{1}{T}\frac{\partial^2M}{\partial X^i\partial
X^j}\qquad
X^i=(S,Q)\nonumber\\
g_{11}=\frac{3Q^2-S}{2S(S-Q^2)},\qquad
&g_{21}&=g_{12}=-\frac{2Q}{S-Q^2},\qquad g_{22}=\frac{4S}{S-Q^2}
\end{eqnarray} we will neglect the fluctuations of  parameter $J$ in all
calculations, which will lead to a wrong flat geometry ($R=0$).
Therefore, the above non-vanishing scalar curvature is a result
coming from another dimension specified by $J$ which fluctuates even
if we set it to zero.  We see that by setting $J=\Lambda=0$, the
metric elements of KN-AdS state space (Eqs. 7-13) reduce to the RN
ones (Eq. 15) but the curvature does not behave in the same manner.
This behavior originated from the existence of an extra dimension in
the parameter space. When the extremal limit is approached, however,
the curvature diverges strongly. Another difference is also observed
from reports in certain works on RN black holes phase space: the
absence of 'Davies phase transition points'. This relates to the
difference between ordinary thermodynamics of extensive systems and
that of black holes, which constitute non-extensive, non-additive
thermodynamic systems because of the well known scaling of the black
hole entropy with area rather than with volume. However, the
interpretation of the divergence in
  specific heats as phase transitions is not settled and has been
  the
subject of much debate
\cite{curir,pavon,kaburaki1,kaburaki2,sorkin}. Ruppeiner
formalism relies on the usual thermodynamic properties of
extensive systems but it  can also  be applied  to black holes
near the divergent points \cite{arcioni}. So we can use the
Ruppeiner method as a probe to find phase transitions.
Furthermore, Ruppeiner proposed that $\abs{R}$ sets the limiting
lower volume in which the classical fluctuation theory provides a
good approximation \cite{ruppeiner3}. For black holes, it could be
obtained  from (7) that even for a non-charged black hole,
Ruppeiner curvature has a $\frac{1}{S}$ remanent. Replacing $S$
by $ \frac{s}{k_B }$, where $s$ is the entropy per volume, and
leaving the choice of natural units, one could derive a lower
volume, $v\simeq\frac{G\hbar r}{c^3}$ (in which $r$ is the
Schwarzchild radius), above which the classical thermodynamic
regime could be used. This also indicates that if the Schwarzchild
radius of the black hole reaches the Planck length,
$l_p=\sqrt{\frac{G\hbar}{c^3}}$,  the classical thermodynamic
description will break down.
\par Still another problem is that of stability.
In order to determine the points where a change of thermodynamic
stability occurs, we use Poincar\'{e} turning point method.
According to Arcioni and Lozano-Tellechea \cite{arcioni}, by
plotting the conjugate variables ($\beta=\frac{1}{T},
\frac{\phi}{T}=\frac{1}{T}\frac{\partial M}{\partial Q}$) in
according to their extensive parameters ($M, Q$) one could see the
points where black hole  changes its stability (Figs. 3a \& 3b).
There is no turning point and, therefore, no changes of stability
are shown along the plot. The first diagram shows  that $\beta$
has a minimum at $M=\frac{2}{\sqrt{3}}$ (by setting $Q=1$). As
described in \cite{arcioni}, this is not a measure of changing
stability, so nothing special happens at this point (Davies
point).
\begin{figure}
\centering \psfig{file=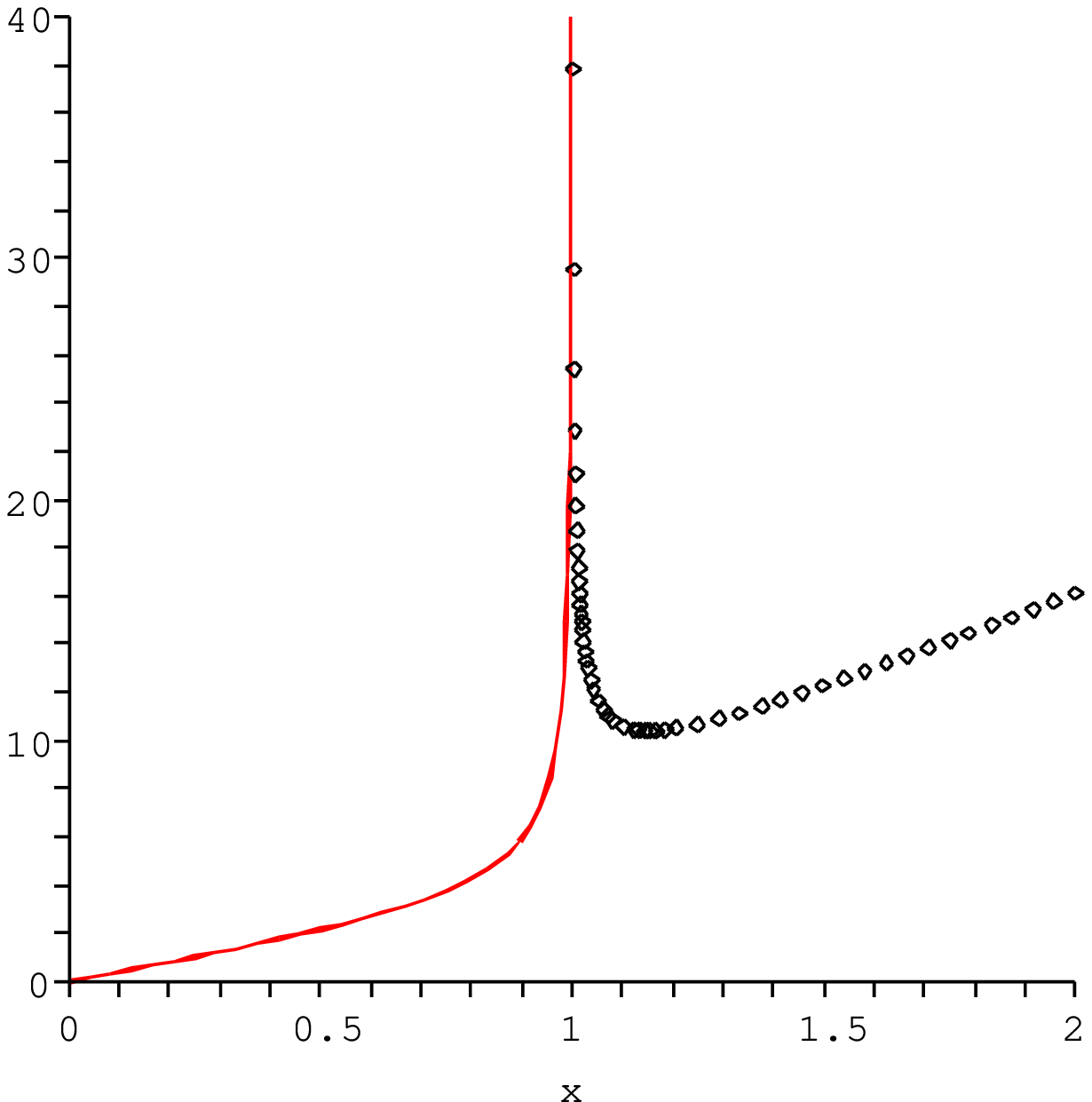, width=.4\textwidth,
height=4cm} 
\caption{ {\footnotesize {\bf a)} Dotted line - $\beta$ as a
function of $x(=M)$ at fixed $Q$ ($Q=1$). $M\rightarrow\infty$ is
the Schwarzchild limit, $M=1$ is the extremal limit and the plot
has a minimum at $M=\frac{2}{\sqrt{3}}$ (Davies point) but there
is no
turning point. \\
 {\bf b)} Line -$\frac{\phi}{T}$ as a function of
$x(=Q)$ at fixed $M$ ($M=1$). $Q\rightarrow 0$ is the Schwarzchild
limit and $Q=1$ is the extremal limit.  } } \label{fig: 4 }
\end{figure}
\section{Kerr black holes}
 Aman {\it et al.} also calculated the Ruppeiner and Weinhold's
 curvatures for  rotating black holes against a  flat background. They
 reported a vanishing Weinhold curvature for Kerr-type black holes
 but also noted that the physical meaning of this property was not
 clear which they referred to as the {\it ad hoc} definition of Weinhold
 geometry \cite{aman1, aman2}. Using the definition of Weinhold's metric
 \cite{weinhold} for the complete phase space of parameters (KN-AdS black hole),
 and taking the limits  $Q\rightarrow0$, $l\rightarrow\infty$, a non vanishing Weinhold
curvature is derived. The phase transition points could be
obtained by calculating the Ruppeiner curvature
\be{R_{kerr}=\frac{S(S^2+36J^2)}{S^4-16J^4}} $R_{kerr}$ diverges
at the extremal limits: $J=\pm\frac{S}{2}$ (Fig. 4). Further, one
could easily check the instabilities by using the Poincar\'{e}
diagrams: $(\beta(M),M)$ and $(\frac{\omega(J)}{T},J)$. The plots
do not show any turning points (and thus no instability) as
indicated before in \cite{kaburaki1,kaburaki2,arcioni}.
\section{Geometry of  Quasilocal Thermodynamics }
 In this section we replace our definition of the internal energy
of black holes by quasilocal energy proposed by Brown and York
\cite{brown1,brown2} which is derived from the Hamiltonian of
spatially bounded gravitational systems.
 \par The Brown-York derivation of the
quasilocal energy, as applied to a four-dimensional (4D) spacetime
solution of Einstein gravity can be summarized as follows. The
system one considers is a 3D spatial hypersurface $\Sigma$ bounded
by a 2D spatial surface $B$ in a spacetime region that can be
decomposed as a product of a 3D hypersurface and a real
\begin{figure}
\centering \psfig{file=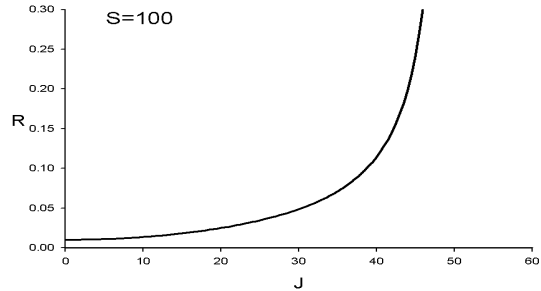, width=.4\textwidth, height=4cm}
\caption{ {\footnotesize Ruppeiner scalar of Kerr BH as a function
of $J$. $R$ diverges   at two values of $J$.  Even at $J=0$, $R$ has
a $\frac{1}{S}$ remnant.  } } \label{fig: 5 }
\end{figure}
line-interval representing time. The time-evolution of the boundary
$B$ is the surface $^3B$. One can then obtain a surface
stress-tensor on $^3B$ by taking the functional derivative of the
action with respect to the 3D metric on $^3B$. The energy surface
density is the projection of the surface stress-tensor normal to a
family of spacelike surfaces such as $B$ that foliate $^3B$. The
integral of the energy surface density over such a boundary $B$ is
the quasilocal energy associated with a spacelike hypersurface $S$
whose orthogonal intersection with $^3B$ is the boundary $B$. It is
assumed that there are no inner boundaries, such that the spatial
hypersurfaces $S$ are complete. In the case where horizons form, one
simply evolves the spacetime inside as well as outside the horizon.
Under these conditions, the QLE (Quasi Local Energy) is defined as:
\be{E=\frac{1}{8\pi}\oint_B d^2x\sqrt{\sigma(k-k_0)}} where $\sigma$
is the determinant of the 2-metric on $B$, $k$ is the trace of the
extrinsic curvature of $B$, and $k_0$ is a reference term that is
used to normalize the energy with respect to a reference spacetime,
not necessarily flat. To compute the QLE for asymptotically flat
solutions, one can choose the reference spacetime to be flat as
well. In that case, $k_0$ is the trace of the extrinsic curvature of
a two-dimensional surface embedded in flat spacetime, such that it
is isometric to $B$.
\par For spacetimes that are asymptotically flat
in spacelike directions, the quasilocal energy and angular
momentum defined there agree with the results of Arnowitt, Deser
and Misner in the limit that the boundary tends to spatial
infinity \cite{adm, martinez,bose}. We have used ADM parameters
in the above discussions. Now, ADM mass ($M$) is replaced by
quasilocal energy of KN black holes
\be{E=r_0[1-\sqrt{1-\frac{2M}{r_0}+\frac{a^2+Q^2}{r_0^2}}]+O(\frac{a^2}{r_0})}
where $a=\frac{J}{M}$ and $r_0$ is the radius of the bounding
surface. Here we only consider the small electrical charge and
slow rotating regime $(\frac{|a|}{r_0}\ll 1),$ so the definition
of $Q$ and $J$ do not have to change. Inserting
\be{M=\sqrt{\frac{S}{4}+\frac{1}{S}(J^2+\frac{Q^4}{4})+\frac{Q^2}{2}}}
in (18) one could derive a new fundamental equation for quasilocal
parameters
\be{E(S,Q,J)=r_0\bigl[1-\sqrt{1-\frac{2\sqrt{\frac{S}{4}+\frac{1}{S}(J^2+\frac{Q^4}{4})+\frac{Q^2}{2}}}{r_0}+\frac{a^2+Q^2}{r_0^2}}\bigr]+O(\frac{a^2}{r_0})}
Now the Ruppeiner metric components could be calculated
\be{g_{ij}=\frac{1}{T}\frac{\partial^2E}{\partial X^i\partial
X^j}\qquad X^i=(S,Q,J)} For RN black holes (2-dimensional state
space)  curvature is a function of $r_0$,  such that  \be{\lim_{
r_0\rightarrow\infty} R_{RN}(r_0,S,Q)=0} For Kerr-Newmann  black
holes (full thermodynamic state space) the calculated curvature is
different from $R_{RN}$ for any finite value of $r_0$ (they are
too large to be written here) and has the following limiting form:

\be{\lim_{r_0\rightarrow\infty,
J\rightarrow0}R_{KN}(r_0,S,Q,J)=\frac{S^2+Q^2S+2Q^4}{(S^2-Q^4)(S+Q^2)}}
As expected this limit equals to (14). The above considerations
show that our results do not contingent upon computing all
thermodynamic quantities at infinity.
\section{Conclusion}
\par Usual statistical mechanics, augmented by renormalization group
theory, are used to build up from the microscopic physics and deduce
information about the macroscopic world. For the black holes, the
microscopic description needs a quantum theory of gravity which is
still missing.  In contrast,  the covariant thermodynamic
fluctuation theory builds down from the macroscopic equation of
states using the Riemannian geometry \cite{ruppeiner3}. In using
this method one must be careful to consider all possible physical
fluctuations because neglecting one parameter may lead to inadequate
information about the model. For the RN case, the Ruppeiner
curvature could be obtained using the complete set of physical
fluctuating parameters. The resulting Ruppeiner curvature is not
zero,   behaves like other types of black holes, and could be used
to set a lower bound, $R_{Sch}\simeq l_p$,  beyond which the
classical thermodynamic description breaks down as expected. The
Method proposed here could also be applied to other cases, e.g., as
shown by Henneaux and Teitelboim \cite{henax}, it is possible to
promote the cosmological constant to a thermodynamic state variable
which may be the subject of  future study.

\end{document}